\documentclass[a4paper,fleqn,usenatbib]{mnras}

\def\msun{M$_\odot$}

\def\kms{km s$^{-1}$~}

\usepackage{newtxtext,newtxmath}

\usepackage[T1]{fontenc}
\usepackage{ae,aecompl}

\usepackage{graphicx}	
\usepackage{amsmath}	
\usepackage{amssymb}	

\title[A globular cluster in Sextans A]{An old, metal-poor globular cluster in Sextans A and the metallicity floor of globular cluster systems}

\author[M.A.Beasley et al.]{
  Michael A. Beasley$^{1,2}$\thanks{E-mail: beasley@iac.es},
Ryan Leaman$^{3}$,
Carme Gallart$^{1,2}$,  
S{\o}ren S. Larsen$^{4}$,
\newauthor
Giuseppina Battaglia$^{1,2}$,
Matteo Monelli$^{1,2}$,
Mario H. Pedreros$^{5}$
\\
$^{1}$Instituto de Astrofisica de Canarias, Calle V\'ia L\'actea, La Laguna, Tenerife, Spain\\
$^{2}$University of La Laguna. Avda. Astrof\'isico Fco. S\'anchez, La Laguna, Tenerife, Spain\\
$^{3}$Max Planck Institute for Astronomy, K\"onigstuhl 17, 69117, Heidelberg, Germany\\
$^{4}$Department of Astrophysics / IMAPP, Radboud University, PO Box 9010, 6500 GL Nijmegen, The Netherlands\\
$^{5}$Departamento de F\'isica, Facultad de Ciencias, Universidad de Tarapac\'a, Arica, Chile\\
}

\date{Accepted XXX. Received YYY; in original form ZZZ}

\pubyear{2019}

\begin{document}
\label{firstpage}
\pagerange{\pageref{firstpage}--\pageref{lastpage}}
\maketitle

\begin{abstract}
  We report the confirmation of an old, metal-poor globular cluster in the nearby
  dwarf irregular galaxy Sextans A, the first globular cluster known in this galaxy.
  The cluster, which we designate as Sextans A-GC1, lies some 4.4 arcminutes ($\sim1.8$ kpc)
  to the SW of the galaxy centre and clearly resolves into stars in sub-arcsecond seeing ground-based imaging.
  We measure an integrated magnitude $V=18.04$, corresponding to an absolute magnitude, $M_{V,0} = -7.85$. This gives an inferred mass  $M\sim$1.6$\times10^5~$\msun, assuming a Kroupa IMF.
  An integrated spectrum of Sextans A-GC1 reveals a heliocentric radial velocity
  $v_{\rm helio}=305\pm15$~\kms, consistent with the systemic velocity of Sextans A. 
The location of candidate red giant branch stars in the cluster, and stellar population analyses of  the cluster's  integrated optical spectrum, suggests a metallicity [Fe/H] $\sim$--2.4, and an age  $\sim9$ Gyr. We measure a half light radius, $R_h = 7.6\pm0.2$ pc. Normalising to the galaxy integrated magnitude, we obtain a $V$-band specific frequency, $S_N=2.1$.  We compile a sample of 1,928 GCs in 28 galaxies with spectroscopic metallicities and find that the low metallicity of Sextans A-GC1 is close to a "metallicity floor" at [Fe/H] $\sim$--2.5 seen in these  globular cluster systems which include the Milky Way, M31, M87 and the Large Magellanic Cloud. This metallicity floor appears to  hold across 6 dex in host galaxy stellar mass and is seen in galaxies with and without  accreted GC subpopulations.
\end{abstract}




\section{Introduction}

The continuing discovery of globular clusters (GCs) in Local Group galaxies indicates the census
of star clusters in these systems is incomplete (e.g., Georgiev et al. 2009; Hwang et al. 2011; Veljanoski et al. 2013; Huxor et al. 2014; Mackey et al. 2016; Caldwell et al. 2017).
For example, Cole et al. (2017)  "re-discovered"   a massive  GC (first identified by Hoessel \& Mould 1982) in the central regions
of the Pegasus dwarf irregular galaxy (DDO216), making it the lowest-luminosity Local Group galaxy presently
known to host a massive (log (M/\msun)$ > 4.0 $) GC. Similarly, Wang et al. (2019)
have confirmed the existence of a  sixth star (and perhaps globular) cluster (first noted by Shapley 1939) in the Fornax dwarf spheroidal,  which appears to be in the latter stages of tidal disruption.

The observational characterisation of the GC systems of dwarf galaxies is particularly important in order to tackle
a number of astrophysical issues. For example, GCs are generally known to be old systems thought
to form during periods of intense star formation (Beasley et al. 2002, Kravtsov \& Gnedin 2005, Brodie \& Strader 2006; Pfeffer et al. 2018). Therefore, the identification of old GCs in galaxies
is indicative of at least one early, intense phase of star formation (e.g., Caldwell et al. 2017).
This brings insight into the earliest phases of galaxy formation and is
something that can be contrasted against resolved star formation histories in nearby galaxies (Gallart et al. 2015). 
In addition, the dynamical state of a GC system can be used to infer properties
of the mass distribution of the parent galaxy.
Read et al. (2012) have argued that the present configuration
of the GCs associated with the Fornax dSph likely indicates a weakly-cusped dark matter density
profile in the galaxy (but see also Goerdt et al. 2006 who argue for a core), with the additional requirement that the present Fornax dSph GCs were accreted
at an earlier time. A comparable analysis has been performed for the Pegasus dwarf (Leaman et al., submitted).

Another interesting observation is that the total mass of a GC system appears to  scale linearly with the virial
mass of the host galaxy (Blakeslee et al. 1999; Spitler 2009, Georgiev et al. 2010; Harris et al. 2013, Forbes et al. 2018). This observation
has potentially important implications for the build-up of galaxy haloes and GC systems.
The $M_{\rm GC} -M_{\rm vir}$ relation is likely driven by merging and is a consequence of the central limit theorem in the hierarchical assembly paradigm (El-Badry et al. 2019). Therefore GCs in dwarf galaxies are prime locations to learn about the physics of GC formation efficiency (prior to any merging activity which may confuse the picture). In addition, the relation can be used in principle to infer virial masses for galaxies without resorting to using either the highly non-linear M$_{\rm star}$ - M$_{\rm halo}$  relations or dynamical measurements (Beasley et al. 2016;  Prole et al. 2019).
Also, since the haloes of galaxies are thought be built up by the accretion of satellites,  understanding the inventory of GCs in dwarfs 
opens up the possibility of using GC systems as tracers of the mass accretion histories of galaxies (C\^ot\'e et al. 1998; Boylan-Kolchin et al. 2017; Beasley et al. 2018; Pfeffer et al. 2018).

In this contribution, we confirm the existence, and explore the properties of,  a massive, old GC in the 
dwarf irregular galaxy Sextans A (D = 1.42 Mpc; Bellazzini et al. 2014) via high-resolution ground-based imaging 
and integrated light (IL) spectroscopy. 
Sextans A lies in the NGC 3109 association, which is located  on the  outskirts of (and may or may not be associated with) the Local Group (see e.g., Davidge 2018). 
The association comprises a loose group of galaxies consisting of NGC 3109, the Antlia dwarf, Sextans A, Sextans B and Leo P (Bellazzini et al. 2013). The faint dwarf galaxy Leo B is also thought to belong to this group (Sand et al. 2015).
In terms of GCs, 
Demers et al. (1985) identified 10 candidates in NGC 3109, but as far as we are aware these have not been confirmed.
In addition, a conference proceeding by
Blecha (1988) suggests some 23 unconfirmed candidate GCs in this  galaxy.
Georgiev et al. (2008) looked for old GCs in the Antlia dwarf galaxy using archival HST imaging
and found no obvious candidates. Sextans B is presently known to host at least one  young ($\sim2$ Gyr),
massive star cluster (Sharina et al. 2007; see also Bellazzini et al. 2014) but no known  GCs.  Neither Leo B nor Leo P are presently thought to host GCs.

The plan of this paper is as follows: In Section~\ref{Data} we describe the imaging and spectroscopic data
used in this study. In Section~\ref{Analysis}, we describe the identification of the GC and our measurements 
of its physical properties. In Section~\ref{Floor} we focus on the low metallicity of this cluster and compare it to a 
compilation of  spectroscopic metallicity measurements for Galactic and extragalactic GCs. Finally, in  Section~\ref{Conclusions} we summarize  our conclusions.

\section{Data} \label{Data}

\subsection{Las Campanas and Subaru  Imaging}

\begin{figure*}
	\includegraphics[width=12cm]{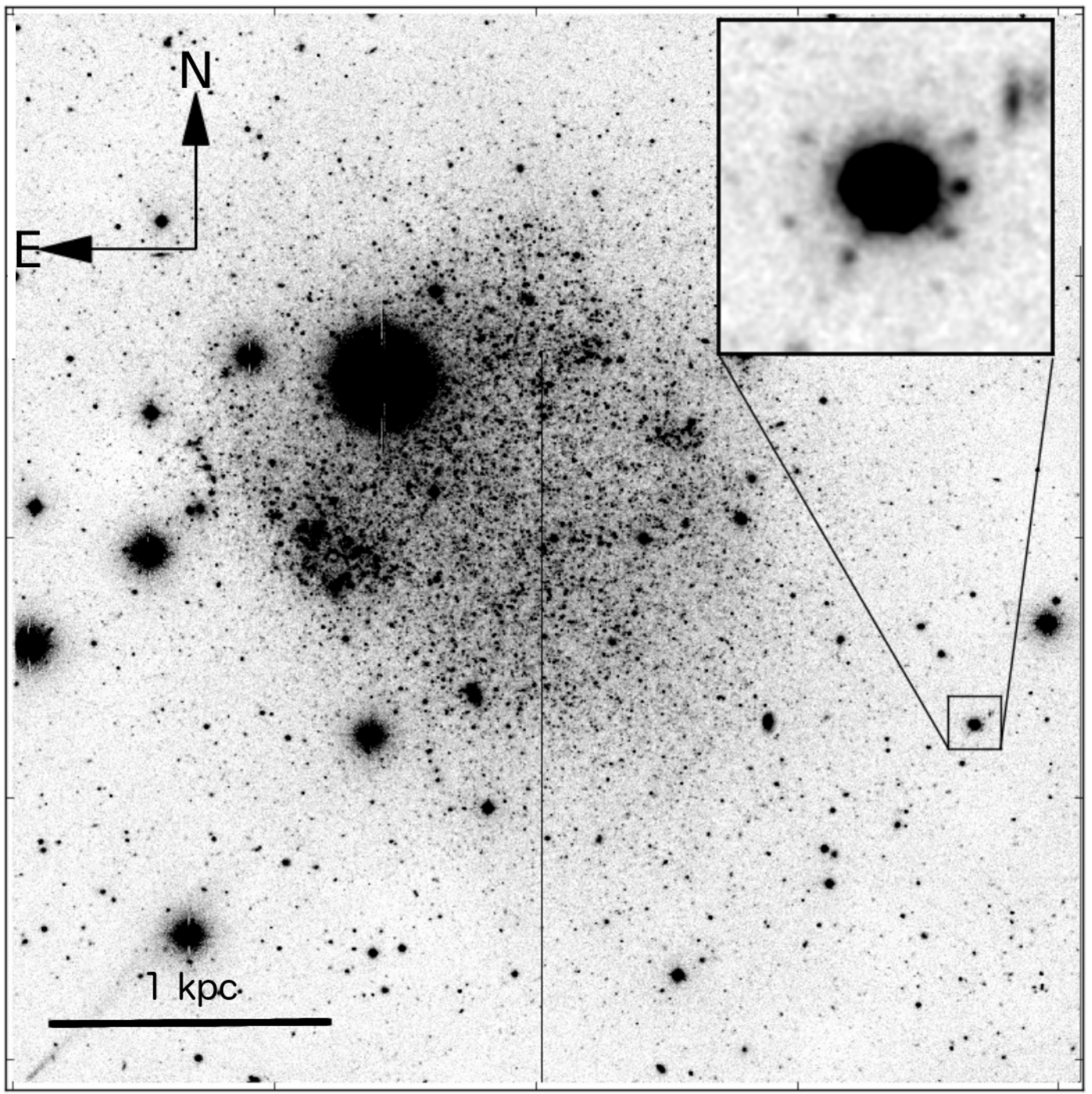}
    \caption{$V$-band image from the  du Pont 100 inch  of Sextans A with the globular cluster Sextans A-GC1 marked.}
    \label{fig:fig1}
\end{figure*}

Pedreros \& Gallart (2002) performed a search for GCs in Sextans A and identified one  candidate (their object  13) which is the subject of this work.  A subset of the imaging analysed here (and by Pedreros \& Gallart 2002) comprises V and I CCD images of Sextans A which were obtained in February 1997 with the Du Pont 100 inch
telescope at the Las Campanas Observatory (see Pedreros \& Gallart 2002). A 2048x2048 Tektronix CCD with a scale
of 0.296 arsec/pixel was used, giving a field of view of 8 x 8 arcminutes.
Three adjacent fields were observed in Sextans A with median seeing of 0.9$\arcsec$.
The central field was centred on Sextans A, a second field centred 7 arcminutes west
of the galaxy centre and the third 7 arcminutes south of the galaxy centre.
Total integration times per field were 2400s in V and 1800s in I.
We also searched for archival imaging of Sextans A taken in good seeing and found V,I,R imaging of the object taken with Subaru Suprime Cam. These data have a 0.2 arsec/pixel scale and were taken in seeing varying between 0.8 and 1.0 arcsec.  Visual inspection of a region some 15 arcminutes in radius ($\sim6$ kpc) in the Subaru  imaging revealed no further, obvious GC candidates. 

The cluster candidate lies some 4.4 arcminutes ($\sim$1.8
 kpc) to the SW of the galaxy centre (Fig~\ref{fig:fig1}).  Visual inspection of the imaging reveals that its outskirts clearly resolve into stars in sub-arcsecond imaging. The cluster candidate presents an extended, slightly elongated appearance with "crinkly" edges characteristic of semi-resolved star clusters (e.g. Huxor et al. 2014).  As detailed in Pedreros \& Gallart (2002), $V$ and $I$ PSF photometry of the images was obtained using stand alone versions of DAOPHOT/ALLFRAME (Stetson 1987). Total magnitudes for the cluster candidate were obtained for each image and filter using an aperture radius of 8 pixels, or 2.4\arcsec, and then averaged. A relatively small aperture was used in order to minimize random errors in the magnitudes due to contamination by close neighbours in crowded regions. Comparison between total magnitudes obtained with this aperture radius and a larger one of 30 pix (8.9\arcsec) shows that we may be underestimating the magnitudes by up to $\simeq$ 0.5 magnitudes. We measure $V=18.04$, $(V-I)=0.98$, which at our adopted distance to Sextans A, and correcting for reddening, gives an absolute magnitude for the cluster of  $M_{\rm V,0} = - 7.85$. 

\subsection{Spectroscopy}\label{Spectroscopy}

In order to confirm object 13's association with Sextans A, we required a radial velocity.
IL optical spectroscopy of Sextans A-GC1 was obtained with OSIRIS (Cepa et al. 2000)
on the GTC in La Palma on March 5, 2016. We used the 2000B grism with slit width 1 arcsecond and integrated for 600s.  Seeing was 1.0 arcsec and airmass 1.2. The data were reduced (bias-subtraction, flat-fielding, wavelength
calibration and flux-calibration) using a combination of IRAF and Python scripts.
The final spectrum covers a wavelength range $3947-5693$~\AA, has a resolution (FWHM) of $3.0\pm0.2$~\AA, and 
a median signal-to-noise of 27 \AA$^{-1}$. 

\section{Analysis}\label{Analysis}

\subsection{Cluster Identification}

We measure a heliocentric radial velocity for the 
cluster of $v_{\rm helio}=305\pm15$~\kms using FXCOR in IRAF. Within FXCOR  the spectrum was logarithmically
re-binned and cross-correlated with the stellar population models of Vazdekis et al. (2010; 2015). The best cross-correlation
solution was obtained for an old (12 Gyr), metal-poor ([Fe/H]=--2.3) and $\alpha$-enhanced ([$\alpha$/Fe]=0.4) model. 
This is in agreement with a 
more detailed assessment of the stellar population (Section~\ref{Stellarpops}).
The (HI) heliocentric velocity for Sextans A is $324\pm2$~\kms~(Koribalski et al. 2004), therefore Sextans A-GC1 has a heliocentric velocity consistent the Sextans A systemic velocity.
Given object 13's photometric properties and its kinematic association with Sextans A, we identify it as a 
globular cluster and henceforth call it Sextans A-GC1.

\begin{figure}
	\includegraphics[width=8cm]{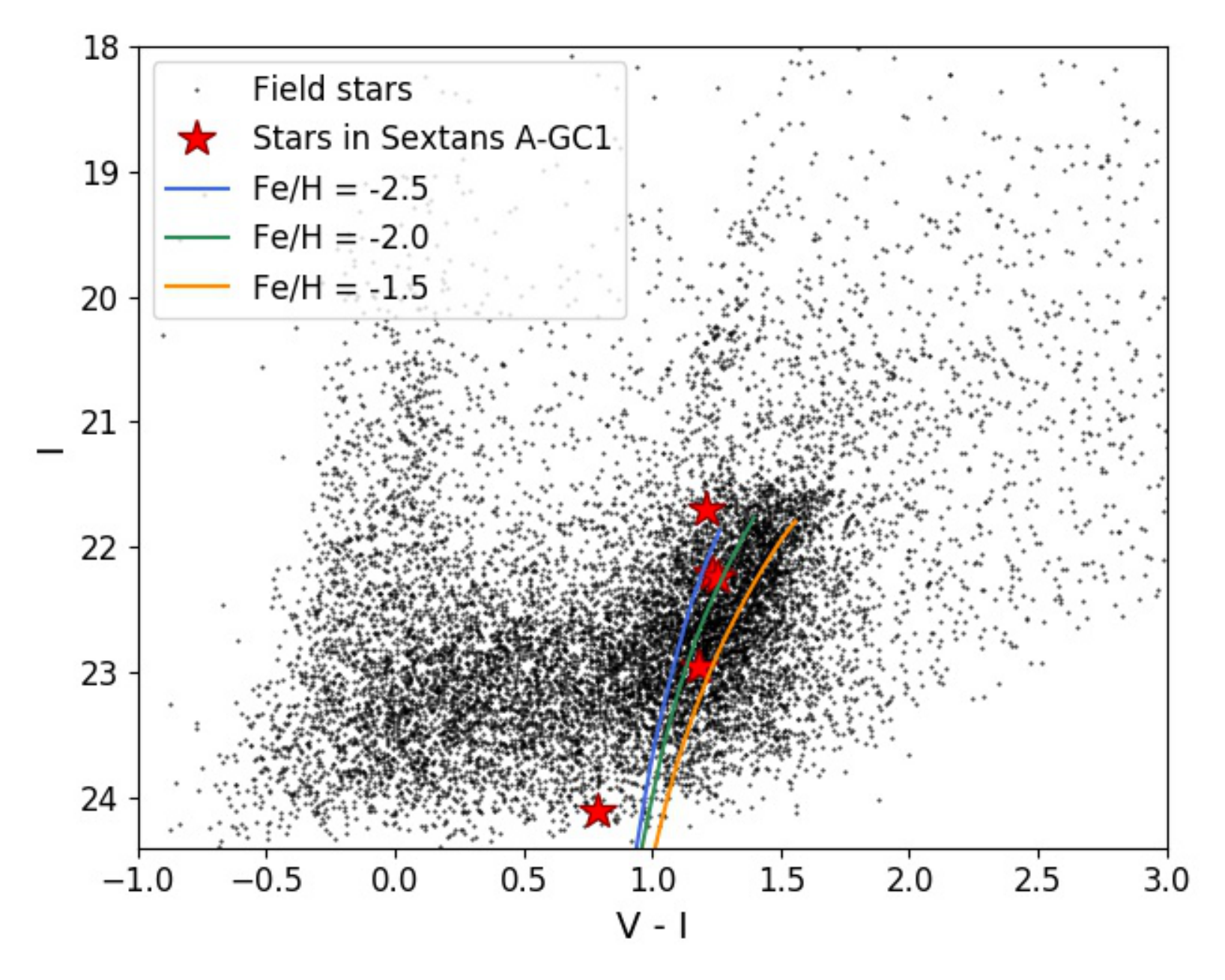}
    \caption{Colour-magnitude diagram for the field stars in Sextans A and candidate giant branch stars in the globular cluster Sextans A-GC1 constructed from the du Pont imaging. Overplotted are isochrones from Dotter et al. (2016) with age 12 Gyr, [$\alpha$/Fe]=0.4 for three different metallicites.}
    \label{fig:fig2}
\end{figure}

The quality of the Las Campanas imaging is such that we can measure magnitudes for a few candidate resolved upper red giant branch (RGB) stars in Sextans  A-GC1 and plot them in a colour-magnitude diagram. This is shown in Fig~\ref{fig:fig2}, where we compare five candidate cluster RGB stars (identified by eye) to photometry of the field stars in Sextans A. The stars that we identify as RGB stars in the CMD  are visible  in the inset of the image of the cluster shown in Fig~\ref{fig:fig2}. Although only suggestive, and offering no strong age information, the presence of an  RGB would imply a stellar population older than about 2 Gyr. Assuming an old age, the colour of the RGB suggests  a low metallicity. Using the Dotter et al. (2016) isochrones, adopting an age of 12 Gyr (appropriate for Milky Way GCs) and [$\alpha$/Fe]=0.4 (Section~\ref{Stellarpops}), we infer a  metallicity for the cluster in the range  $-2.5<$[Fe/H]$<-2.0$ . A younger cluster would imply a somewhat more metal-rich stellar population (for 3 Gyr isochroness, [Fe/H]$\sim-1.5$).

\subsection{Size and ellipticity}\label{Size}

Shape parameters for Sextans A-GC1 were obtained  using iShape (Larsen 1999) measured from the 
du Pont $V,I$ and Subaru $V,R,I$  images.
We fit a series of PSF-convolved profiles using PSFs determined with the IRAF version of DAOPHOT. Inspection of the iShape residuals indicated that a MOFFAT (EFF25 in iShape) profile best matched the cluster light profile (this produced the lowest $\chi^2$). We measure an ellipticity, $\epsilon=0.12\pm0.01$ and a circularised half-light radius of $1.10\pm0.03$ arcsec, corresponding to $R_h = 7.6\pm0.2$ pc for D = 1.42 Mpc. 
Uncertainties on the parameters were obtained by taking the standard deviation of the ellipticities and sizes
between the  images of the two datasets.

\begin{figure}
	\includegraphics[width=8.0cm]{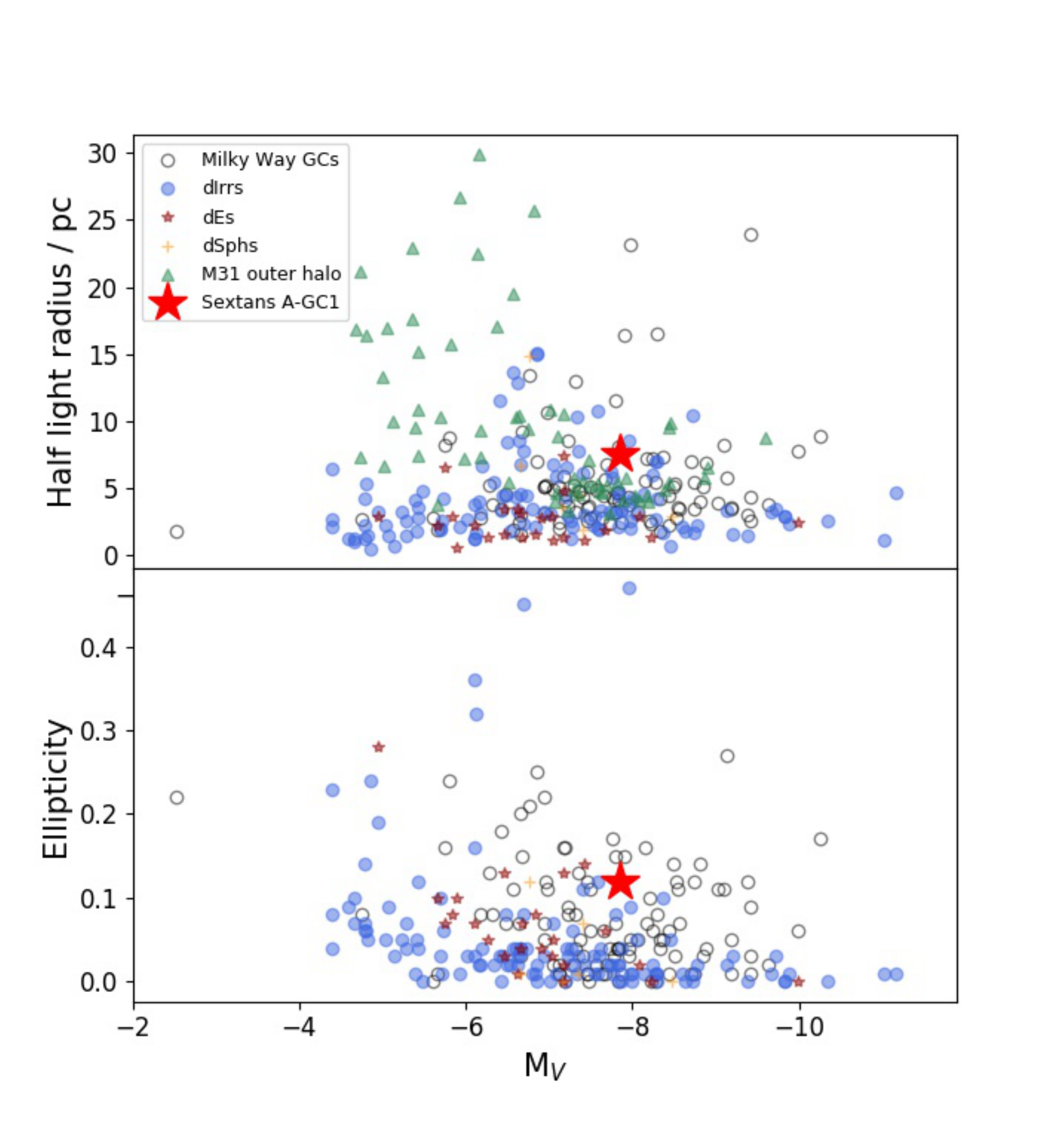}
        \caption{{\it Top panel:} Half light radius of Sextans A-GC1 compared to globular clusters in the MW, M31 and nearby dwarfs. {\it Bottom panel:} Ellipticity of Sextans A-GC1 compared to globular clusters in MW, M31 and nearby dwarfs.}
    \label{fig:fig3}
\end{figure}

In Fig~\ref{fig:fig3} the size and ellipticity of Sextans A-GC1 is compared  to GCs in the Milky Way (Harris 1996), M31 (Huxor et al. 2014) and  a variety of nearby dwarf galaxies (Georgiev et al. 2009). 
In terms of half-light radius, the Sextans A cluster is quite large.
With $R_h = 7.6\pm0.2$ pc, it is somewhat larger than the median
sizes for GCs in the Milky Way ($R_h=4.4$, $\sigma=3.9$ pc)
and nearby dIrrs ($R_h=4.4$, $\sigma=3.9$ pc), and is closer to the median for M31 "outer halo" GCs ($R_h=7.3$, $\sigma=6.5$ pc). Sextans A-GC1 is also quite elliptical ($\epsilon=0.12\pm0.01$), which  is higher than the median  Milky Way ($0.06$, $\sigma=0.07$) and dIrr ($\epsilon=0.03\pm0.08$).

\subsection{Stellar population properties of Sextans A-GC1}
\label{Stellarpops}

\begin{figure}
	\includegraphics[width=9cm]{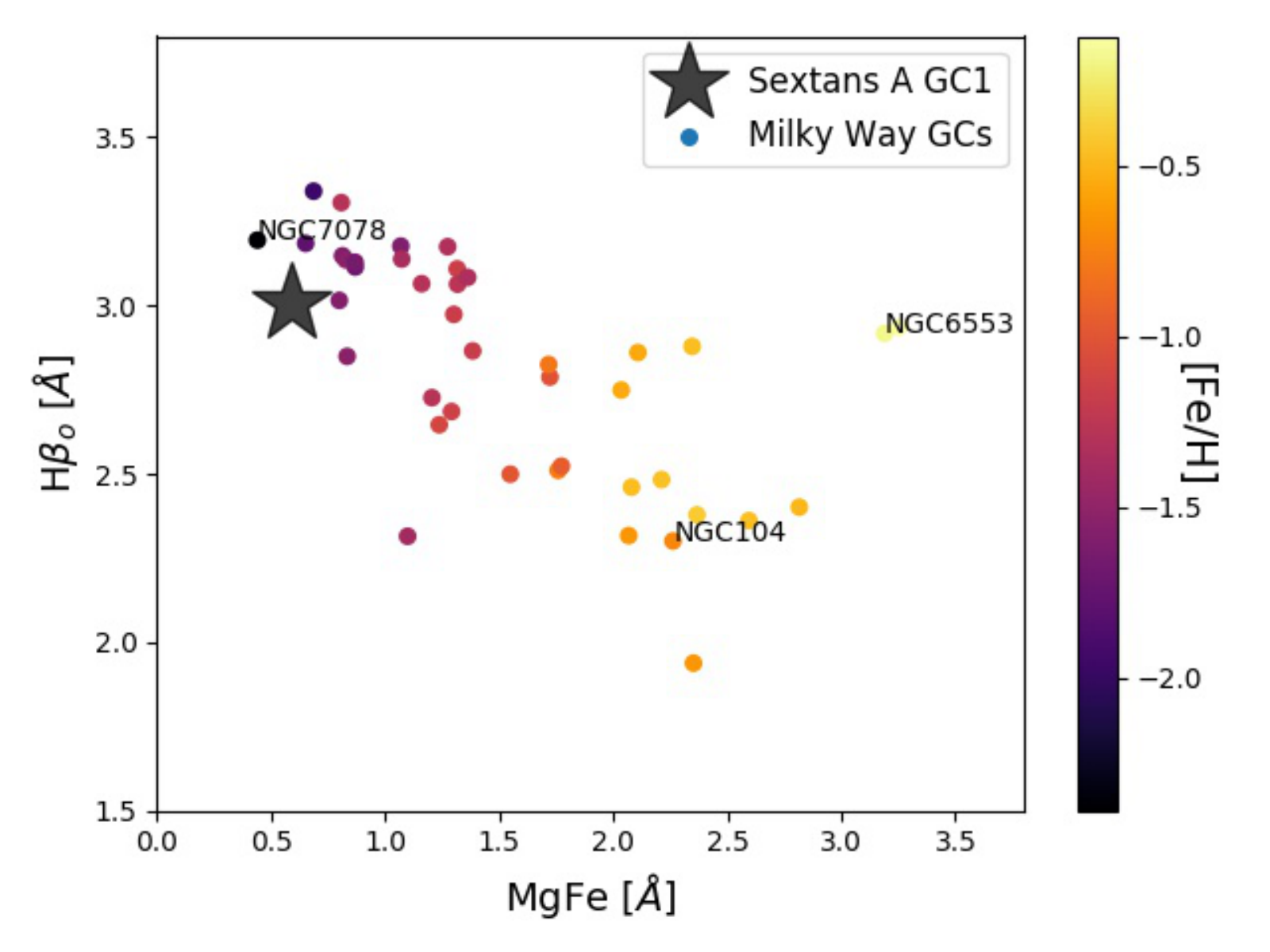}
        \caption{Location of Sextans A-GC1 compared to Milky Way globular clusters (Schiavon et al. 2005) with metallicities from the compilation of Roediger et al. (2014). The cluster lies in a region consistent with it being old and metal-poor.}
    \label{fig:fig4}
\end{figure}

Figure~\ref{fig:fig4} compares metallicity-sensitive ([MgFe])
and age-sensitive (H$\beta_{o}$; Cervantes \& Vazdekis 2009) indices of Sextans A-GC1 with Milky Way 
GC data using the spectra of Schiavon et al. (2005).
For the purposes of the comparison all the spectra have been
convolved to a common spectral resolution of 3.5~\AA.
The Milky Way GCs have been colour coded by their metallicities taken from the compilation of Roediger et al (2014).
The figure indicates that Sextans A-GC1 lies in a region which is old and metal-poor, and is located close to the metal-poor ([Fe/H] = --2.5) Galactic globular cluster M15.  The apparent bifurcation of the Milky Way GCs into two sequences in H$\beta_o$ is probably real, and appears to reflect a differing specific  fraction of blue straggler stars in the central regions of GCs (Cenarro et al. 2008). In this sense, Sextans A-GC1 may be associated with those GCs with a relatively lower  fraction of blue stragglers contributing  to the integrated light spectrum.  We also note that the distribution of stars on the horizontal branch can also effect the Balmer lines (e.g. Beasley et al. 2002; Schiavon et al. 2004; Perina et al. 2011).

\begin{figure}
 
	\includegraphics[width=8cm]{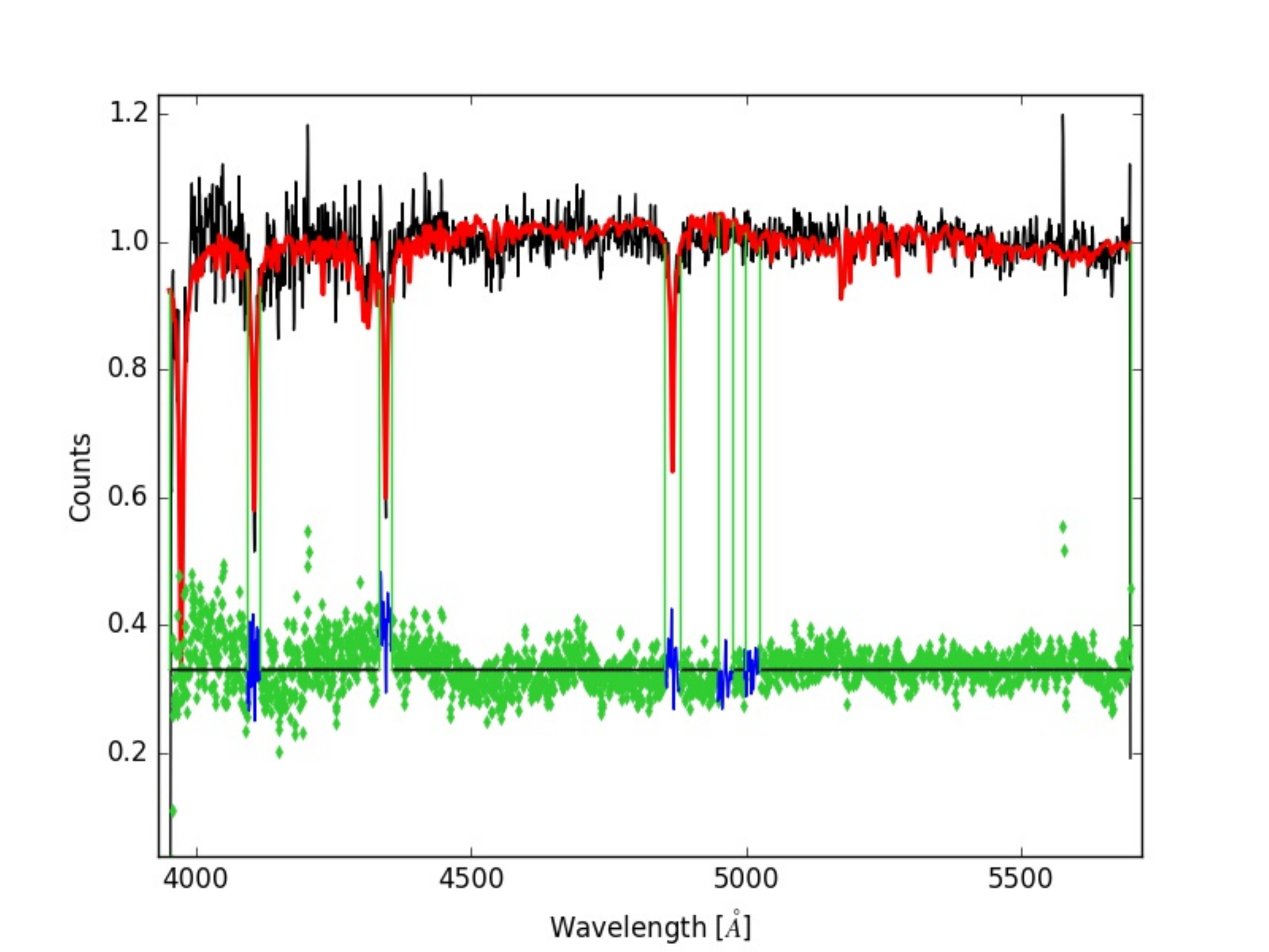}
        \caption{Best-fitting pPXF combination of MILES model templates (red) compared to the GTC spectrum of
          Sextans A-GC1 (black). Residuals from the fits are shown in green, pixels rejected from the fits are shown in blue. }
    \label{fig:fig5}
\end{figure}

\begin{figure}
	\includegraphics[width=9cm]{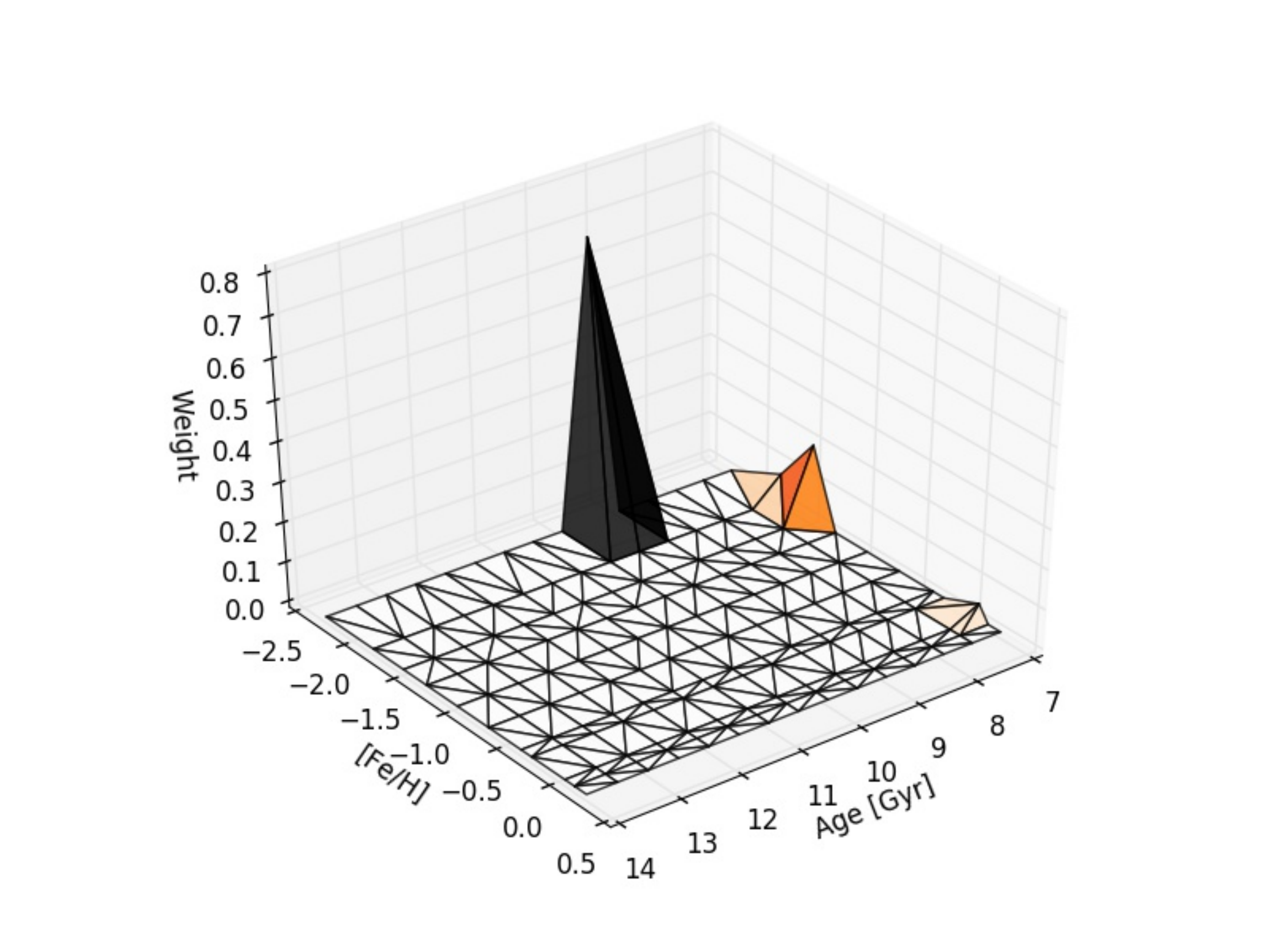}
        \caption{Best age and metallicity solutions from the pPXF fits. The favoured solution is old and metal-poor. }
    \label{fig:fig6}
\end{figure}

To determine stellar population parameters for the GC, we used pPXF (Cappellari \& Emsellem 2004). This provided the best-matching combination of templates based on the MILES models to give age, metallicity and [$\alpha$/Fe] estimates (Fig.~\ref{fig:fig5} and Fig.~\ref{fig:fig6}). We applied a multiplicative polynomial of order 10 to correct the continuum shape, and used a regularization  based on the recipe given in the pPXF documentation. We also apodised the leading and trailing 20 pixels of the spectrum to avoid edge-effects. 
The optimal template gives age = $8.6\pm2.7$ Gyr, [Fe/H] = $-2.38\pm0.29$ and [$\alpha$/Fe] = $0.29\pm0.18$.  
Small fractions of "younger" solutions (Fig.~\ref{fig:fig6})
probably reflect hot stellar populations not properly accounted
for in the models (Ocvirk et al. 2006).

Estimation of the uncertainties in this 
analysis is not straightforward; experiments showed that Monte Carlo simulations which simply add artificial noise to the spectra tend to significantly underestimate the true uncertainties in pPXF. Therefore, the uncertainties given above reflect the standard deviations of each parameter for various solutions which include varying the polynomial for continuum normalisation and the wavelength range covered in PPXF. We note that [$\alpha$/Fe] is particularly problematic given the low metallicity of the cluster. The inferred age  is sensitive to the  horizontal branch morphology of the cluster (the MILES models assume a canonical mass-loss formula on the HB; Vazdekis et al. 2010), however this effect is to some extent mitigated by performing a full-spectral fit as opposed to relying on the temperature-sensitive Balmer lines alone.

For the age and metallicity of the cluster, the Vazdekis et al. (2010) models predict a $V$-band mass-to-light ratio, $\Upsilon_{V}$=1.38 (Kroupa IMF). For an absolute magnitude $M_{V,0} = -7.85$, this gives an inferred mass,  $M\sim$1.6$\times10^5~$\msun. Bellazzini et al. (2014) quote an absolute magnitude for Sextans A,  $M_{V,0} = -14.2$. This implies a GC specific frequency, $S_N=$2.1.

\begin{table}
	\centering
	\caption{Basic and derived parameters for Sextans A GC1}
	\label{tab:table1}
	\begin{tabular}{ll} 
		\hline
		R.A.(J2000) & 10:10:43.80 \\
		Dec.(J2000) & --04:43:28.8 \\
		V & 18.04 \\
		V-I & 0.98\\
		  Rh (pc) & $7.6\pm0.2$\\
		  $\epsilon$ & $0.12\pm0.01$\\
         RV (kms) & $305\pm15$\\
         Fe/H & $-2.38\pm0.29$\\
         Age (Gyr) & $8.6\pm2.7$\\
		\hline
	\end{tabular}
\end{table}

\begin{table*}
	\centering
	\caption{Various statistical properties of the metallicity distributions of globular cluster systems}
	\label{tab:table2}
	\begin{tabular}{lccccccl} 
	\hline
	\hline
	Galaxy & Median [Fe/H] & Q1 & Q3 & IQR & Min [Fe/H]$^{*}$ & Max. [Fe/H]$^{**}$ & Source\\
		\hline
		SLUGGS  & $-0.64$ & $-1.21$ & $-0.21$ & $1.0$ & $-2.60$ & $1.13$ & Usher et al. (2012)\\
		M87 & $-0.95$ & $-1.29$ & $-0.68$ & $0.61$ & $-2.24$ & $0.11$& Cohen, Blakeslee \& Ryzhov (1998) \\
		Cen A & $-0.82$ & $ -1.26$ & $-0.45$ & $ 0.81$ & $ -2.15$ & $ 0.0$ & Beasley et al. (2008)\\
		Sombrero & $-1.13$ & $-1.51$ & $-0.65$& $0.86$& $-2.80$& $0.38$& Alves-Brito et al (2011)\\
		M31 & $-1.00$ & $ -1.40$ & $ -0.60$ & $ 0.80$ & $-2.50$ & $0.40$ & Caldwell et al. (2011)\\
		Milky Way & $-1.32$ & $-1.70$ & $-0.78$ & 0.92 & $-2.48$ & $0.0$ & Harris (1996) \\
		M33 & $-1.24$ & $-1.55$ &$-0.93$ &$0.62$ &$-1.74$ &$-0.83$ & Larsen et al. (2018); Beasley et al. (2015)\\
		VCC1087 & $-1.51$ & $-1.76$ & $-1.34$ & $0.42$ & $-1.88$ & $-1.15$ & Beasley et al. (2006)\\
		LMC & $-1.72$ & $-1.84$ & $-1.55$ & $0.29$ & $-2.02$ & $-1.30$ & Piatti et al. (2018); Beasley et al. (2002b)\\
		SMC & $-1.28$ &&&&$-1.28$&$-1.28$& Dalessandro et al. (2016)\\
		NGC 205 & $-1.10$ & $-1.20$ & $-0.80$ & $0.40$ & $-1.30$ & $-0.60$ & Sharina et al. (2006)\\
		Sextans A & $-2.38$ & &&&$-2.38$&$-2.38$& This work\\
		NGC 147 & $-2.06$ & $-2.37$ & $-2.01$ & $0.36$  & $-2.48$ & $-1.54$ & Larsen et al. (2018)\\
		NGC 6822 & $-2.00$ & $-2.43$ & $-1.67$ & $0.76$ & $-2.53$ &  $-1.27$ & Larsen et al. (2018); Hwang et al. (2014)\\
		NGC 185 & $-1.50$ & $-1.63$ &  $-1.45$ & $0.18$ & $-1.68$ & $-1.42$ & Sharina et al. (2006)\\
		WLM & $-1.95$ & &&&$-1.95$&$-1.95$& Larsen et al. (2014)\\
		Fornax dSph & $-2.10$ & $-2.30$ & $-2.10$ & $0.20$ & $-2.50$ & $-2.10$ & Larsen et al. (2012); Strader et al. (2003)\\
		Pegasus dIrr & $-2.05$ & &&&$-2.05$&$-2.05$& Leaman et al., submitted\\
		\hline
	\end{tabular}
	
	$^{*}$ minimum datapoint with metallicity > Q1 - 1.5$\times$IQR. $^{**}$ maximum datapoint with metallicity < Q3 +1.5$\times$IQR
\end{table*}

\section{Discussion} \label{Floor}

\subsection{The metallicity floor of globular clusters}

The metallicity of Sextans A-GC1 is comparable to the  most metal-poor GCs in the Milky Way. This in itself is unsurprising. It has been known since Brodie \& Huchra (1991) that the mean spectroscopic metallicity of a globular cluster system scales with the mass (stellar and total) of the host galaxy. Sextans A, being a relatively low mass system ($M_{*}\sim4.4\times10^7$~\msun; McConnachie 2012) might be expected to host low metallicity clusters.

However, given the very low metallicity of Sextans A-GC1 ([Fe/H]$\sim-2.4$), it is interesting to ask: what is the minimum (and maximum) metallicity 
that GCs possess, and what does this tell us about their formation compared to the host galaxy field stars (see Carney 1996, Forbes et al. 2018 and Kruijssen 2019 for related discussions)?
Stars with [Fe/H]$\sim-3.0$ are often termed extremely metal-poor (EMP) stars (Beers \& Christlieb 2005); are there any extremely metal-poor GCs (EMPGCs) in galaxies? Is there a minimum metallicity for GC formation?

\begin{figure*}
	\includegraphics[width=18cm]{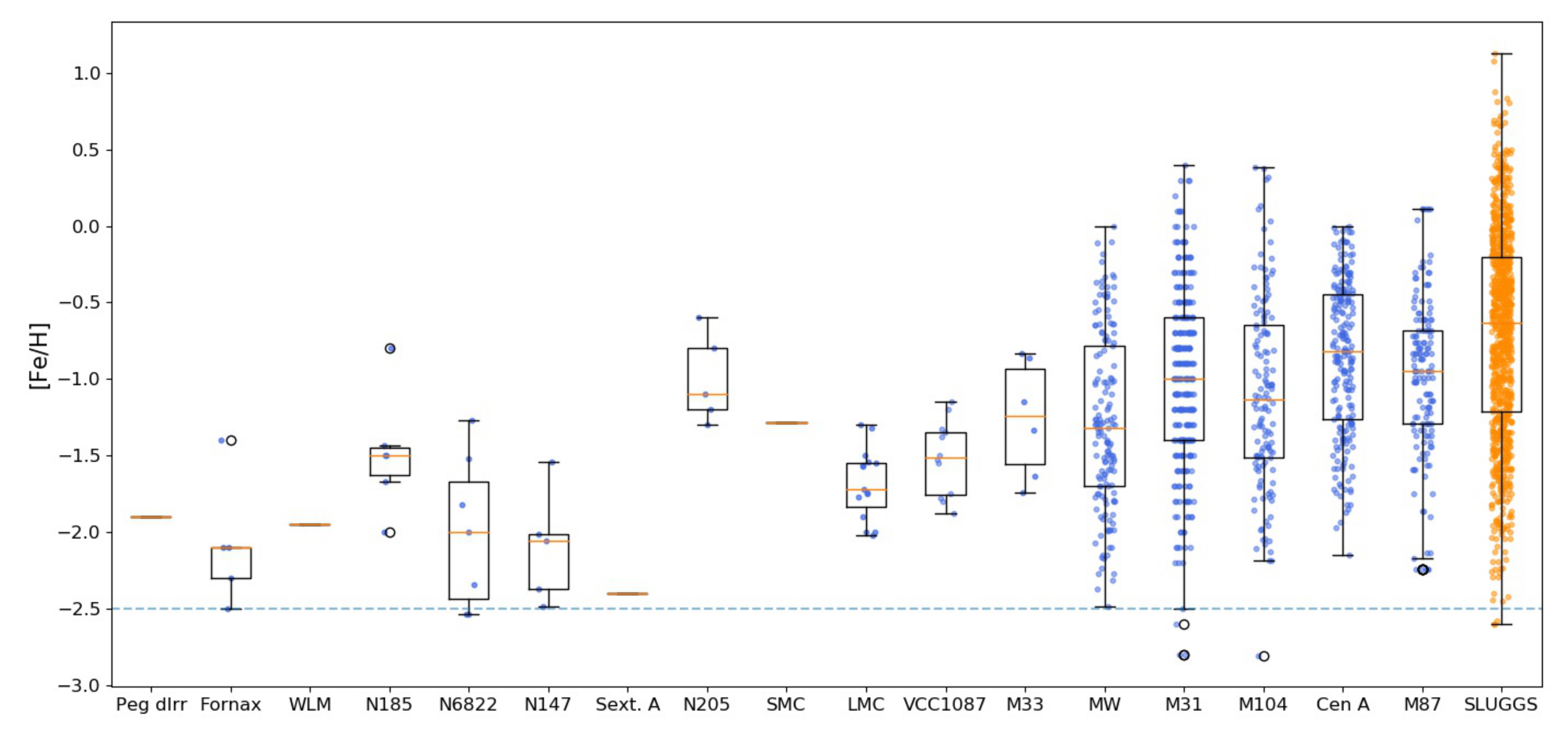}
        \caption{Box-and-whisker plots of the metalllicity distributions of globular cluster systems of galaxies with spectroscopic metallicities (1,928 GCs). Galaxies have been ordered by increasing stellar mass. Blue points show data for individual galaxies. The SLUGGS composite galaxy data is shown in orange. Single horizontal orange lines indicate GC systems with only one known GC. We have added jitter in the $x$-axis to make the datapoints more readily visible.
}
    \label{fig:fig7}
\end{figure*}

In Fig.~\ref{fig:fig7} we show box-and-whisker plots for a compilation of 1,928 GCs in 28 galaxies with spectroscopic metallicity measurements.  Of the 28 galaxies, 11 come from  SLUGGS (Usher et al. 2012). We have combined these  to create a composite GC metallicity distribution to reflect  potentially significant differences in metallicity scale employed (using the NIR calcium triplet, rather than optical spectroscopy). We have excluded studies for individual galaxies which  sample a only a tiny faction of the total GC system (e.g., Larsen et al. 2002; Puzia et al. 2005; Cenarro et al. 2007).
The derived statistical properties and data sources for the globular cluster systems are given in Table 2. 
Where necessary we have converted between total metallicity, [M/H], and [Fe/H] using [Fe/H] = [M/H] - A[$\alpha$/Fe], with  A = 0.75 (see Vazdekis et al. 2015). In the cases where  [$\alpha$/Fe] is unavailable we have assumed [$\alpha$/Fe] = 0.3 which is reasonable for old GCs (Roediger et al. 2014). The SLUGGS [Z/H] metallicities have been converted to [Fe/H] using eqn. 2 of Usher et al. (2012).

Caveats here are important. Most surveys of massive galaxies focus on the brightest GCs, which might provide a biased view of the true GC  metallicity distribution. This is particularly true of galaxies which exhibit a clear "blue-tilt" in their colour-magnitude distributions (Harris et al. 2006; Spitler et al. 2006; Strader et al. 2006) which would bias the metallicities of the bright GCs to higher values.  In addition, despite significant observational efforts, the [$\alpha$/Fe] ratios of extragalactic GCs determined from low-resolution IL spectroscopy are not particularly  well constrained. This uncertainty impacts directly on the conversion between total metallicity ([Z/H]) and the iron metallicity ([Fe/H]). It is also not clear what the minimum metallicity low-resolution IL techniques are capable of measuring. As metallicity decreases, the metal-lines become progressively weaker to the point that little metallicity sensitivity may remain.  Low-resolution IL techniques work well for metallicities to [Fe/H]$\sim-2.5$ (e.g., for the Milky Way GC M15) but below this value such approaches have not been tested.
Also, importantly,  different methodologies have been employed to derive metallicities of the GCs (either low-resolution IL spectroscopy, high-resolution IL spectroscopy, or high-resolution stellar spectroscopy). However  in the cases where the same objects have been observed with different approaches, the uncertainties are not more that 0.3 dex and likely closer to 0.2 dex (see e.g., Colucci et al. 2013; Larsen et al. 2018).  

Fig.~\ref{fig:fig7} reproduces the general trend that more massive galaxies have, on average, more metal-rich GC systems (Brodie \& Huchra 1991; Strader et al. 2004; Peng et al. 2006).
The upper bound metallicity lies at around solar metallicity for the most massive (non-SLUGGS) galaxies with well-studied globular cluster systems. Combining the Milky Way, M31, Sombrero, NGC 5128 and M87 GC systems, only 22/961 GCs ($\sim$2\%) have [Fe/H]$>$0.0. This differs from  the stars in the central regions of massive galaxies which  are generally found to have super-solar  metallicity (e.g., Trager et al. 2000; Gallazzi et al. 2005; Martin-Navarro et al. 2018)\footnote{We also note that the central regions of massive early-type galaxies show evidence for variable  stellar initial mass functions (bottom-heavy IMF; i.e., dominated by low mass stars; La Barbera et al. 2013) which raises the question of whether some of these clusters have been born with such an IMF.}.
Examining the SLUGGS sample, 130/902 ($\sim$14\%) have [Fe/H]$>$0.0. This relatively large fraction of high-metallicity GCs may be real, but given the uncertain behaviour of the CaT at high metallicities (e.g. Vazdekis et al. 2003), which has not been calibrated above solar metallicity (Usher et al. 2012), we believe these metallicities may be overestimated.

Interestingly, Fig.~\ref{fig:fig7} indicates that there seems to be a relatively clear lower bound for the GC metallicities, which lies at [Fe/H]$\sim-2.5$. This, for example, is approximately the  metallicity of the well studied metal-poor Galactic GC M15 (age = 12.75 Gyr; Vandenberg et al. 2013).  In the whole sample of 1,928 GCs, only 6 GCs potentially  have [Fe/H]$<-2.5$.  These include two GCs in M31 and one GC in Sombrero with relatively poor-quality spectra (compared to their respective samples). The remaining three clusters lie in the SLUGGS galaxies NGC 3377 (2 GCs) and NGC 4278. 

Inspection of Fig.~\ref{fig:fig7}  also suggests that this "metallicity floor" may be independent of host galaxy mass. Between the Local Group Pegasus dIrr and M87, the cD in the Virgo cluster,  there is a difference of $\sim$6 dex in stellar mass and $\sim$5 dex in virial mass (Zhu et al. 2014; Leaman et al. 2019). However, both systems have similar minimum metallicities in their GCs.
By extension, this lower limit on the GC metallicities may also be  independent of the accretion history of the host galaxy as well. For example, the mass accretion history of Centaurus A (NGC 5128) is likely quite complex (as judged from the metallicity distribution function (MDF) of its GCs; Beasley et al. 2008) yet it exhibits a very similar lower bound on the GC metallicities as dwarf systems such as the LMC or the WLM galaxy.

The origin of the  metallicity floor in GCs is presently unclear. It could be  purely a sampling effect; galaxies have small numbers of EMP stars and therefore have a correspondingly small number of EMPGCs. 
With decreasing dwarf galaxy mass, the fraction of stars below [Fe/H]=$-2.5$ should increase (given the galaxy mass--metallicity relation), and one might expect these systems to be the ones where candidate EMPGCs would be found.  However it may be that these most metal poor galaxies are simply not massive enough to support formation of any clusters large enough to survive to the present day. Clearly, a full census of the GC populations of nearby dwarf systems is required to investigate this point.

In massive galaxies, the precise  metallicity distribution and fraction of metal-poor stars is not known. However, analyses of integrated quantities (Maraston \& Thomas 2000), resolved upper giant branch studies (Rejkuba et al. 2005; Lee \& Jang 2016),  and chemical evolution modelling (Vazdekis et al.  1997;  Pipino \& Matteucci 2004) suggests that the fraction of stars with [Fe/H]$<-2.5$ is probably less than 1\%.
Therefore, assuming a fixed ratio of GC formation with respect to field stars we would expect to observe a similar fraction of  EMPGCs in these galaxies (for reference, 1\% of the M87 GCs system would correspond to $\sim$100 GCs in this galaxy). 
However, observations indicate that the ratio of GCs to stars in galaxies  increases with decreasing  metallicity. This is true both between galaxies (Strader \& Brodie 2006), and within individual galaxies where the fraction of metal-poor GCs 
increases with radius (Lee, Kim \& Geisler 1998; McLaughlin 1999). This implies  that  for massive galaxies, whose haloes and GC systems are thought built-up by the accretion of low-mass satellites (C\^ot\'e, Marzke \& West 1998; Tonini 2013; Beasley et al. 2018), metal-poor  GCs are "over-represented" with respect to field stars. That is, in order to find the most metal-poor GCs, the haloes of massive galaxies might be a good place to look. Given that there appear to be few, if any, EMPGCs in the haloes of giant galaxies (Fig.~\ref{fig:fig7}) (haloes which presumably comprise substantial quantities of accreted dwarf systems) the lack of EMPGCs may indeed be a real, rather than observational, effect.

Another possibility is that a sufficiently metal-poor ISM is simply {\it unable} to form massive bound clusters. This may be, for example, tied to the fragmentation properties of metal-poor gas in a similar fashion to that thought to give rise to  the top-heavy initial function expected for population III stars (e.g., Abel, Bryan \& Norman 2002), or to the inefficiency of cooling at very low metallicities (e.g., Loeb \& Rasio 1994). Or, it may simply be that the ISM enriches  sufficiently quickly that there is insufficient time to form EMPGCs - essentially a G-dwarf problem for GCs. Further exploration of these ideas is beyond the scope of this paper.

\section{Summary and Conclusions}\label{Conclusions}

We have confirmed the existence  of a massive GC in the nearby 
dwarf irregular galaxy Sextans A. The GC lies 4.4 arcminutes ($\sim$1.8~kpc) to the southwest of the galaxy centre, has $M_{V,0} = -7.85$, and we estimate a (photometric) mass of $M\sim$1.6$\times10^5~$\msun. Its relatively close proximity  to the central body of Sextans A suggests that there may well be many more GCs to be discovered in nearby dwarf galaxies. We find that the cluster is quite large ($R_h=7.6\pm0.2$ pc) which puts it in a similar region of parameter space to the  M31 outer halo  globular clusters.

From the integrated light spectrum we find that the  cluster is old and that its metallicity is very low ([Fe/H]$=-2.38\pm0.29$), comparable to the most metal-poor GCs in the MW. Follow-up high-resolution IL spectroscopy of this cluster would be very useful to better constrain its metallicity and obtain individual abundance ratios.

We compile spectroscopic metallicity data for the GC systems of 28 nearby galaxies and identify what appears to be a "metallicity floor" in these GCs which occurs at [Fe/H]$\sim-2.5$.
This floor appears to be independent of host galaxy mass and  galaxy accretion history. We briefly discuss the possible origins of this observed lower limit on GC metallicities, which may be  a size of sample effect,  or perhaps  represents a true physical limit on the metallicity of GCs. To make progress in this area, a spectroscopic campaign to observe the most metal-poor GC candidates in galaxies would be very valuable, in addition to obtaining a more complete picture of the MDFs of nearby galaxy haloes below [Fe/H]$=-2.5$. Understanding this metallicity floor in GC systems may bring valuable insight into the efficiency of star cluster formation, and the early star formation phases in galaxies.

\section*{Acknowledgements}

MAB thanks Ana Chies Santos, Chris Brook and N\'uria Salvador Rusi\~nol for interesting discussions regarding this work. R.L. and S.S.L. acknowledge the Lorentz Center workshop on the Formation of Stars and Massive clusters in Dwarf Galaxies over Cosmic Time for stimulating discussions.
MAB acknowledges support from grant AYA2016-77237-C3-1-P  from  the  Spanish Ministry of Science and Universities (MCIU) and  the Severo Ochoa Excellence scheme (SEV-2015-0548).
G.B.  acknowledges financial support from the Spanish Ministry of Science and Universities  under the Ramon y Cajal Programme (RYC-2012-11537) and the grant AYA2014-56795-P.  CG  acknowledges financial support from the Spanish Ministry of Science and Universities with the grant AYA2017-89076-P.  PyRAF is a product of the Space Telescope Science Institute, which is operated by AURA for NASA.








\appendix



\bsp	
\label{lastpage}
\end{document}